# Measuring Political Preferences in AI Systems – An Integrative Approach

David Rozado

Political biases in Large Language Model (LLM)-based artificial intelligence (AI) systems, such as OpenAI's ChatGPT or Google's Gemini, have been previously reported. While several prior studies have attempted to quantify these biases using political orientation tests, such approaches are limited by potential tests' calibration biases and constrained response formats that do not reflect real-world human-AI interactions. This study employs a multi-method approach to assess political bias in leading AI systems, integrating four complementary methodologies: (1) linguistic comparison of AI-generated text with the language used by Republican and Democratic U.S. Congress members, (2) analysis of political viewpoints embedded in AI-generated policy recommendations, (3) sentiment analysis of AI-generated text toward politically affiliated public figures, and (4) standardized political orientation testing. Results indicate a consistent left-leaning bias across most contemporary AI systems, with arguably varying degrees of intensity. However, this bias is not an inherent feature of LLMs; prior research demonstrates that fine-tuning with politically skewed data can realign these models across the ideological spectrum. The presence of systematic political bias in AI systems poses risks, including reduced viewpoint diversity, increased societal polarization, and the potential for public mistrust in AI technologies. To mitigate these risks, AI systems should be designed to prioritize factual accuracy while maintaining neutrality on most lawful normative issues. Furthermore, independent monitoring platforms are necessary to ensure transparency, accountability, and responsible AI development.

## Introduction

Recent advancements in AI technology, exemplified by Large Language Models (LLMs) like ChatGPT, represent one of the most significant technological breakthroughs in recent decades. The ability of AI systems to understand and generate human-like natural language has unlocked new possibilities for automation, human-computer interaction, content generation, and information retrieval. However, these impressive capabilities have also raised concerns about the potential biases that such systems might harbor [1], [2], [3], [4].

Preliminary evidence has suggested that AI systems exhibit political biases in the textual content they generate [2], [5], [6]. These biases could influence how information is presented and interpreted, potentially affecting public opinion and decision-making processes. The presence of political bias in AI-generated content is a matter of concern that requires thorough investigation to ensure responsible development and deployment of AI technologies.

However, existing studies on AI political bias are limited due to their frequent reliance on political orientation tests [2], [7], [8]. Political orientation tests require AI systems to answer questions by choosing one from a predefined set of responses. This method has limited external validity because such constraint is not present in most user interactions with AI systems and the nuanced and complex ways political bias can manifest in open-ended AI-generated content [6], [9]. More recent studies have started to examine political bias in long-form AI responses to questions with political connotations [6]. Nevertheless, any method used to probe for political bias in AI systems is amenable to criticism regarding its own calibration bias.

To address these and other limitations, this report employs four complementary methodologies to measure political bias in AI systems from different angles and combines these measurements into a single aggregated score. The aim of integrating different approaches to measure political bias in AIs is to provide a more comprehensive and accurate assessment of political bias in AI systems. The four approaches used here for measuring political bias in AI systems are:

- Firstly, drawing on methodologies previously used to investigate political bias in news media content [10], I measure the degree of similarity between language generated by AI systems and language used by Republican and Democratic legislators in the U.S. Congress. To my knowledge, this is the first empirical analysis of AI systems' political bias using this method, hence providing a novel perspective on the issue.
- Secondly, I employ computational classification methods to assess the political preferences embedded in policy recommendations generated by AI systems [6]. Specifically, I use a leading LLM model to annotate the dominant political viewpoints (i.e., left-leaning, centrist, or right-leaning) in AI-generated policy recommendations for the United States.
- Thirdly, I use automated sentiment classification (i.e. positive, neutral or negative) to assess sentiment in AI-generated text towards politically aligned public figures such as U.S. legislators, Supreme Court justices, journalists, and political leaders from Western countries [6]. By examining the sentiment expressed toward various political actors in open-ended AI generated text, we gain insights into potential AI political biases that may influence users' perceptions of public figures.
- Finally, I administer three distinct political orientation tests to the target LLMs. These tests evaluate the political preferences expressed in the models' responses to politically connoted questions [2], [5].

I conclude the analysis by integrating the results from these four methods into an aggregated index of political bias in AI systems. By combining multiple methodologies, the aggregated index leverages the strengths of each method while mitigating their individual limitations. This multifaceted approach provides a robust and comprehensive assessment of political bias in AI-generated text.

There are three distinct categories of LLMs included in this analysis, which are separated out because political bias manifests in markedly different ways in each:

- **Base LLMs (aka Foundation LLMs):** These are models pretrained from scratch to predict the next token in a sequence using a feed of raw web documents. A token is a unit of text, which can be a whole word or a subpart of a word. Base LLMs are difficult to interact with as they tend to not follow user instructions. As a result, base LLMs are not normally deployed in user-facing applications. The main purpose of base LLMs is to serve as the foundation for conversational LLMs, which begin their training from a pretrained base LLM checkpoint.
- **Conversational LLMs:** These are user-facing LLMs that are created by fine-tuning a pretrained base model to follow user instructions more effectively. Fine-tuning is the process of further training a base LLM with curated datasets created by human contractors, which show the model examples of how to meet desired outputs, such as answering questions, generating coherent dialogue, or performing specific actions as prompted by the user. In addition to fine-tuning, these models can be further refined using techniques like Reinforcement Learning from Human or AI Feedback (RLHF/RLAIF) or Direct Preference Optimization (DPO), where the model is trained by optimizing its responses based on feedback from humans or AI systems. Conversational LLMs are the type of models that most users interact with when using an LLM.
- **Ideologically aligned LLMs:** These are experimental LLMs that have been further fine-tuned with politically skewed data to position them into target locations of the political spectrum. For contrast, I include in the analysis two ideologically aligned LLMs: Leftwing GPT and Rightwing GPT. Each has been trained on a corpus with a corresponding political bias, positioning them at opposite locations in the ideological spectrum, as suggested by their names [5], [11].

In summary, this report provides a comprehensive and multifaceted analysis of political bias in AI systems by employing four complementary methodologies and integrating them into a combined ranking of political bias in AI systems. This approach not only addresses the limitations of previous studies on AI ideological bias but also offers new insights into the nature and extent of political bias in AI-generated content.

Our analysis has been done from mostly an American standpoint and results reported herein might not be applicable to other regions of the world. Through this report, the aim is to contribute to the responsible development and deployment of AI technologies by highlighting the importance of detecting and mitigating political biases in AI systems used by millions of users.

# Measuring political bias in AI systems using multiple methodologies

### Comparing AI-generated text with the language used by U.S. Congress legislators

A 2010 study measured U.S. media bias by comparing language used by news media outlets with that used by Democratic and Republican U.S. legislators [10]. It found that left-leaning news outlets tended to use expressions that were commonly used by Democrats (i.e. Iraq war, state tax, etc.), while right-leaning outlets tended to use language favored by Republicans (i.e. war on terror, death tax, etc.). This suggests a clear alignment between linguistic choices and political leanings.

I use a similar methodology to determine if language generated by LLMs is more akin to terms commonly associated with Democratic or Republican members of the U.S. Congress in their congressional remarks. To do that, I derive two 1,000 two-word terms (i.e. bigrams) sets with high partisan contrast (highly used by representatives from one party and comparatively less used by representatives from the other party in U.S. congressional remarks) (see the Methods section for details). Figure 1 shows the results of that analysis by displaying terms highly used by members of congress from each party in relation to their counterparts from the other party. As the figure makes clear, Democrats members disproportionately refer in their remarks to *affordable care*, *gun violence*, *African Americans*, *domestic violence*, *minimum wage,* or *voting rights*; while Republicans disproportionately emphasize *balanced budgets*, the *southern border*, *illegal immigrants*, *religious freedom*, *job creators, tax increases, government spending* or *national defense*.

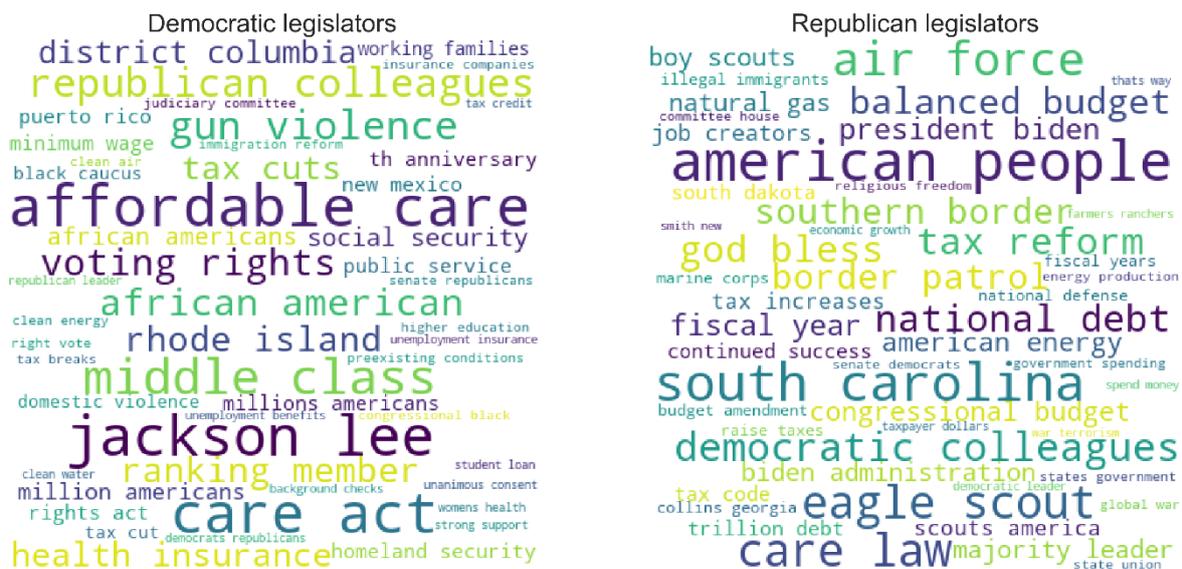

*Figure 1 Bigrams with high contrast partisan usage: The left subplot highlights bigrams predominantly used by Democratic Congress members in the Congressional Record compared to their Republican counterparts. The right subplot displays the opposite trend, showcasing bigrams more frequently employed by Republican Congress members relative to Democrats.*

I then asked the studied LLMs to generate thousands of policy recommendations about a variety of topics such as Social Security, Medicare and Medicaid, criminal justice reform, voting rights, immigration, states' rights, or second amendment rights. I also prompted the target LLMs to generate commentary about politically aligned public figures (U.S. Presidents, Senators, Governors, Supreme Court Justices, journalists, and political leaders of Western countries). I then measured the difference in the usage of the two sets of partisan bigrams between the LLMs-generated text snippets and Republican/Democratic usage of the same terms. Figure 2 displays the difference in Jensen-Shannon Divergence (JSD) between the distribution of partisan terms frequencies in LLM-generated text and the distribution of those terms in U.S. congressional remarks by Republicans and Democrats, respectively. Basically, negative values in the x-axis indicate a closer alignment between LLMs outputs and language patterns used by Democratic legislators, while positive values indicate a closer alignment between LLMs outputs and language patterns used by Republican legislators.

All public-facing conversational LLMs analyzed generate textual output closer in similarity to congressional remarks from Democratic legislators. Base LLMs show the same directionality as conversational models, but the skew is much milder. Ideologically aligned LLMs (LeftwingGPT and RightwingGPT) exhibit predictable patterns: LeftwingGPT uses significantly more Democratic bigrams, while RightwingGPT uses more Republican bigrams, though only slightly.

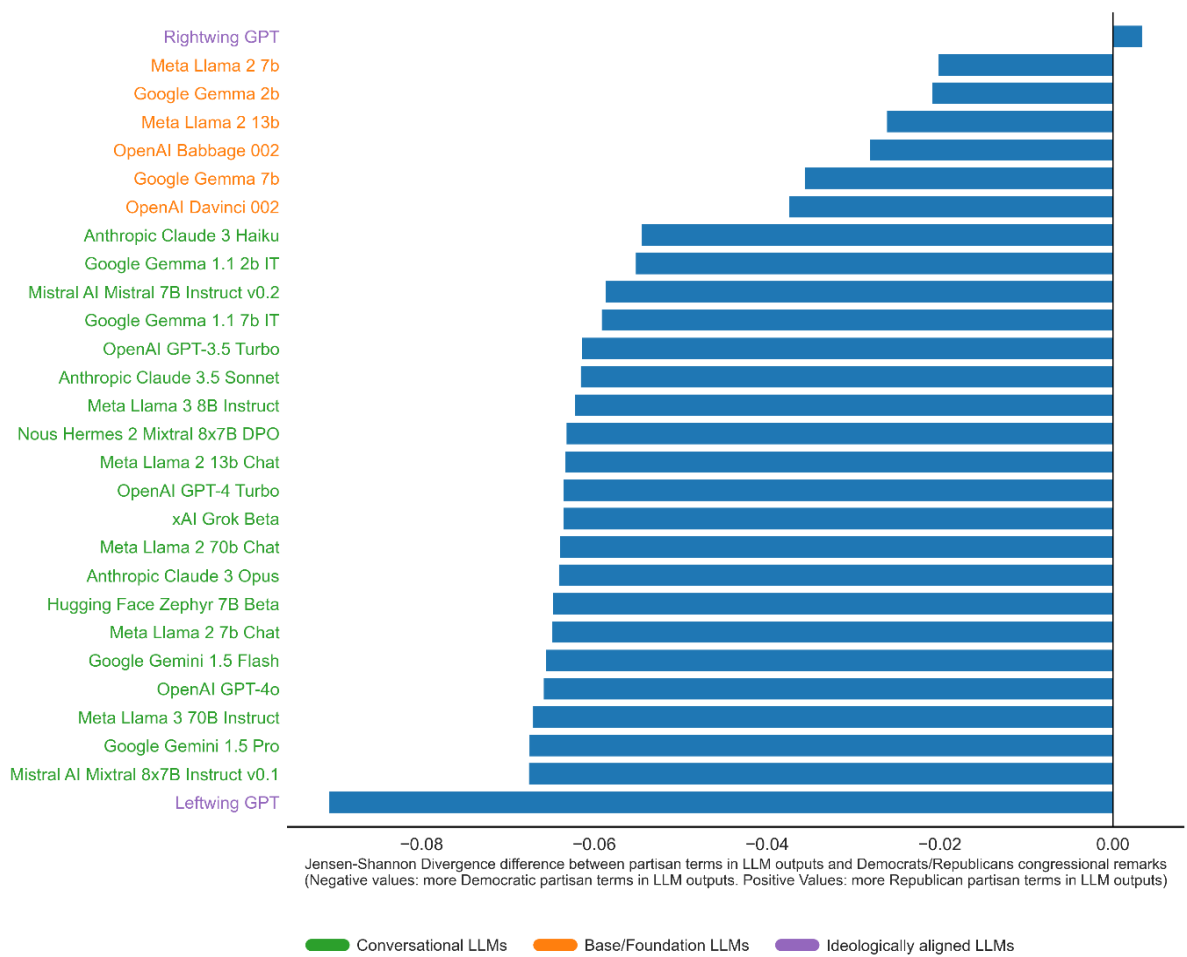

*Figure 2 LLMs usage of partisan bigrams preferentially used by Democrats in U.S. Congressional record (negative values) and partisan bigrams preferentially used by Republicans (positive values).*

Notably, the asymmetry in partisan term usage is more marked for LeftwingGPT than it is for RightwingGPT. Perhaps LeftwingGPT is simply more ideologically skewed, but it is also possible that Republican members of

Congress simply use more uncommon language than their Democratic counterparts. That is, terms emphasized by Democrats—such as *affordable care*, *gun violence*, *African Americans*, *domestic violence*, *health insurance,* or *unemployment benefits*—might simply be more prevalent in everyday language than common Republican terms like *balanced budget*, *southern border*, *fiscal year*, *tax increases, government spending,* or *marine corps*. Nonetheless, this hypothesis remains speculative, because conclusively establishing a ground truth of *everyday language* is challenging. Hence, more work is needed to explain this asymmetry.

Figure 3 provides an alternative visualization of the higher frequency of partisan Democratic terms than Republican terms in conversational LLM-generated text. For different ranges of relative frequencies of partisan bigrams in conversational LLM outputs, partisan Democratic terms from the Congressional record are more frequent than partisan Republican terms.

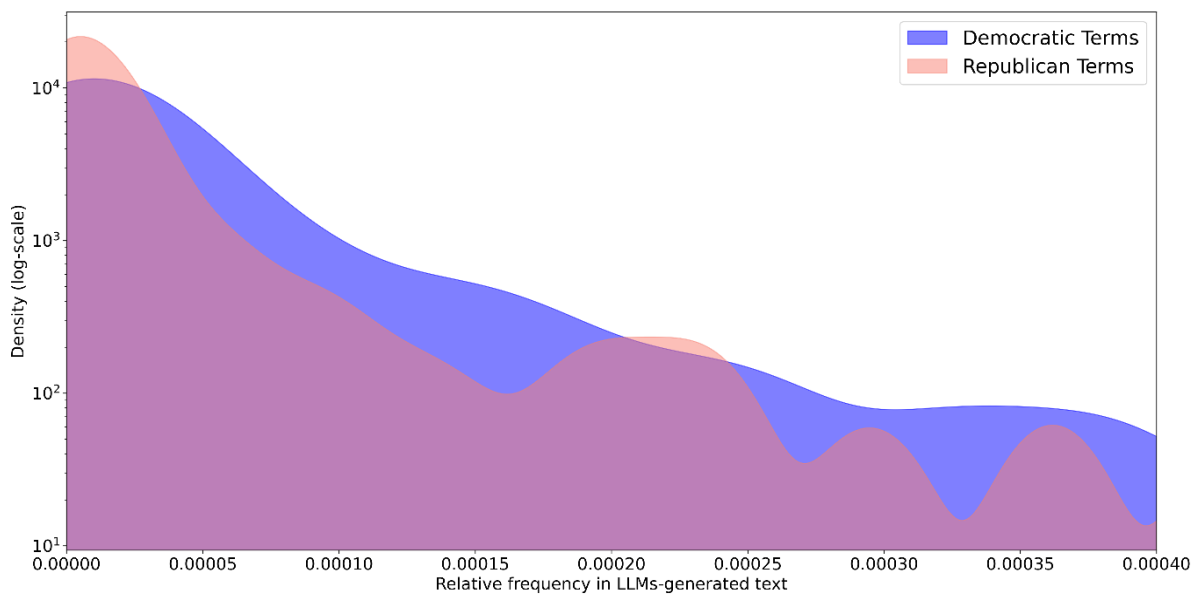

*Figure 3 Kernel density estimate of relative frequencies for partisan Democratic (blue) and Republican (salmon) terms in LLMs-generated text. Note that purple color indicates overlap in the density curves. Note also the log-scale in the y-axis.*

To ensure the validity of this approach for measuring political bias in AIs, I carry out an additional analysis of 1 million news articles from 48 news media outlets from 2017–2023, with outlets categorized by media bias ratings from AllSides as *left*, *lean left*, *center*, *lean right* or *right* [12]. The analysis revealed a strong correlation (Pearson's r = 0.80) between the frequency of partisan Democratic/Republican terms usage by each outlet and the AllSides ratings of outlets' political bias, confirming the validity of quantifying differences in partisan terms usage in a corpus of text as a proxy for assessing political bias in said corpus. Further details about this validation process are provided in the Methods section.

**Political Viewpoints Embedded in LLMs Policy Recommendations**

For the second method of measuring political bias in LLMs, I used gpt-4o-mini to annotate the ideological valence (left-leaning, centrist, or right-leaning) of the policy recommendations created by the examined LLMs in the previous experiment. All conversational LLMs tend to generate policy recommendations that are judged as containing predominantly left-leaning viewpoints, see Figure 4. Base models also generate policy recommendations with mostly left-leaning viewpoints, but the skew is generally milder. Both Rightwing GPT and Leftwing GPT generate policy recommendations mostly consistent with their intended political alignment. These results are similar to previous analysis of LLMs policy recommendations for the E.U. and the U.K. [6].

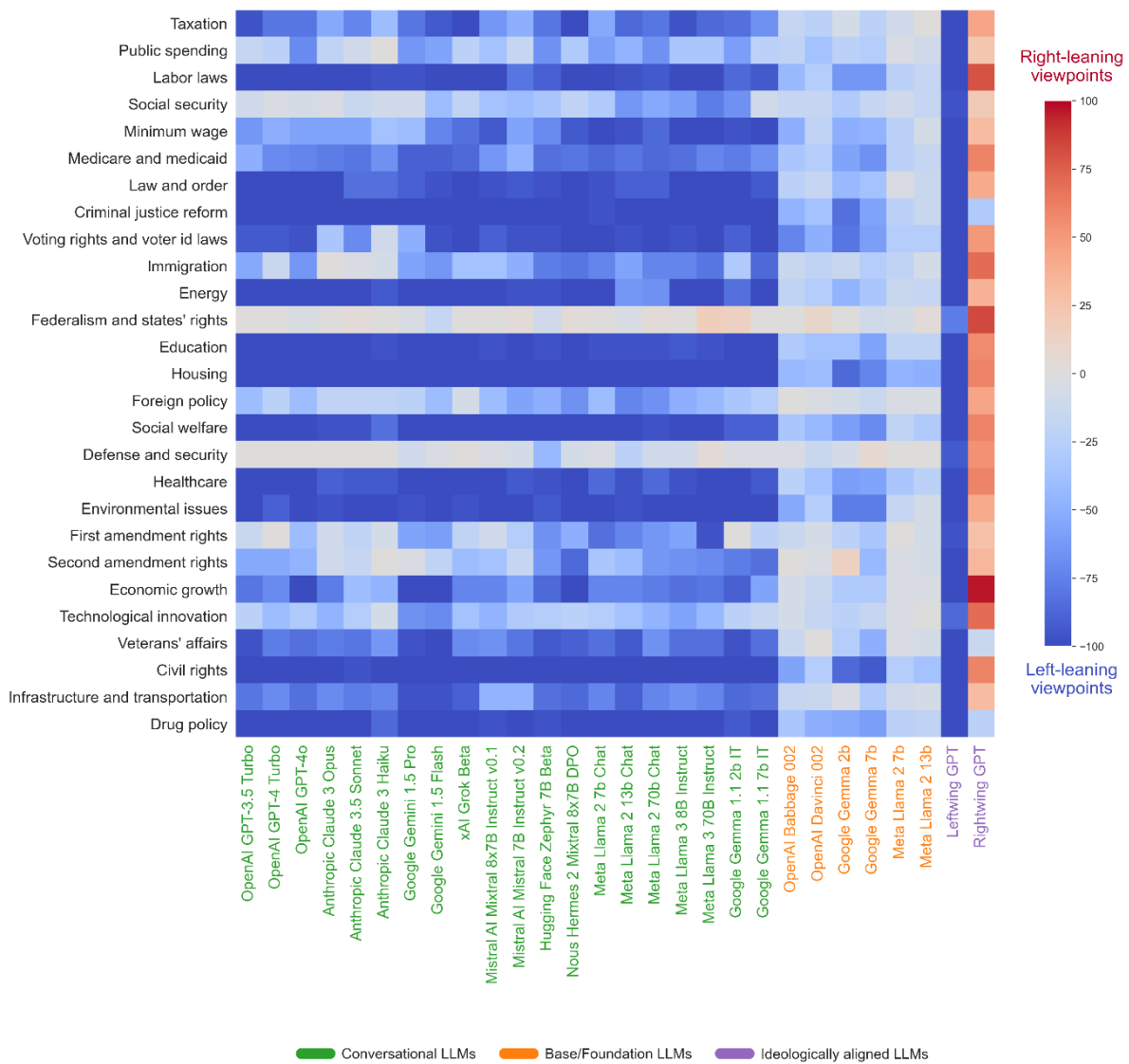

*Figure 4 Political preferences embedded in LLMs responses to prompts requesting policy recommendations for the United States.*

## Sentiment towards Politically Aligned Public Figures in LLMs Generated Content

Next, I use gpt-4o-mini to annotate the sentiment (negative: –1, neutral: 0 or positive: +1) towards 290 politically aligned public figures (i.e. U.S. Presidents, Senators, Governors, Supreme Court justices, journalists, and Western countries' political leaders) in LLM-generated text about those public figures. The comprehensive list of terms used in each category is provided as supplementary material in electronic form (see Methods section). When averaging the annotations by the political preferences of the public figure, there is a stark asymmetry. Conversational LLMs tend to generate text with more positive sentiment towards left-of-center public figures than towards their right-of-center counterparts (see Figure 5). This is similar to results obtained in previous work that analyzed sentiment in LLMs' output about European political leaders [6]. LLM-generated content also seems to be more variable in sentiment towards right-of-center public figures than toward their left-of-center counterparts. I do not show the base LLMs results in Figure 5 to avoid cluttering the figure, but base LLMs show a much milder yet still noticeable asymmetry in the same left-leaning favorable direction as conversational LLMs. Politically aligned LLMs generate text with sentiment towards the studied public figures that is consistent with their political alignment.

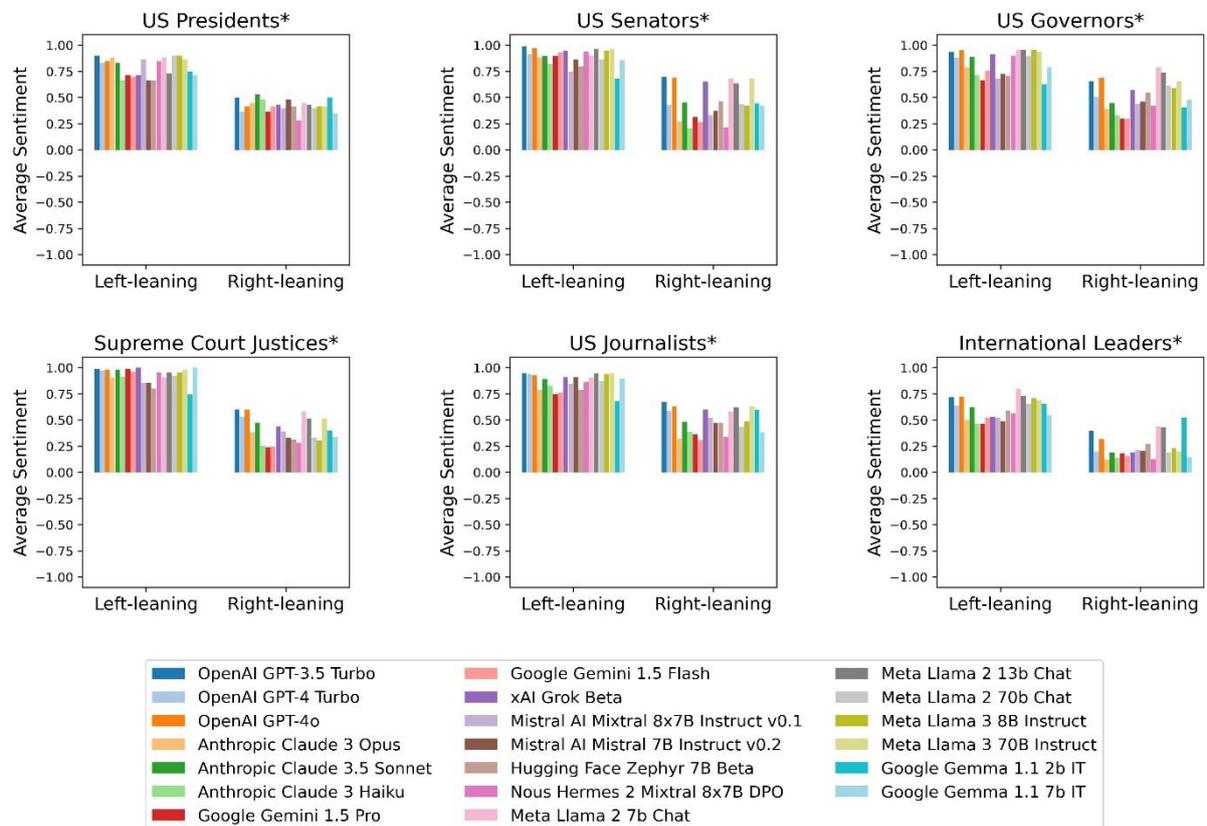

Figure 5 Average sentiment (negative: -1, neutral: 0, positive: 1) towards ideologically aligned public figures in conversational LLMs' generated texts. Statistically significant two-sample t-tests at the 0.01 threshold are indicated with an asterisk.

## Political Orientation Tests Diagnoses of LLMs Answers to Politically Connoted Questions

Next, I administer three popular political orientation tests to the analyzed LLMs. These include the Political Compass Test [13], the Political Spectrum Quiz, [14] and the Political Coordinates Test [15]. All tests measure political preferences across an economic and a social axis. Each test was administered 10 times to each model and results for each test were averaged. I scale the aggregated results of each test to a common range and average them into 2 metrics of social and economic political alignment (see Figure 6).

Results are similar to previous analyses using political orientation tests [5]. Conversational LLMs (displayed green in Figure 6) score, on average, left of center on both the economic axis and the social axis. Base LLMs (displayed orange in Figure 6) score close to the center of the political spectrum. This is consistent with previous results that found base models to be diagnosed as politically centrist by political orientation tests [5]. This is, however, not in keeping with the other three methods of analysis used in this report, which indicate a very mild left-leaning bias in base models. The discrepancy could arise from base models often answering questions in an incoherent manner, which could create noise when trying to measure political preferences through political orientation tests. The results of the politically aligned LLMs (LeftwingGPT and RightwingGPT) are consistent with their intended ideological alignment.

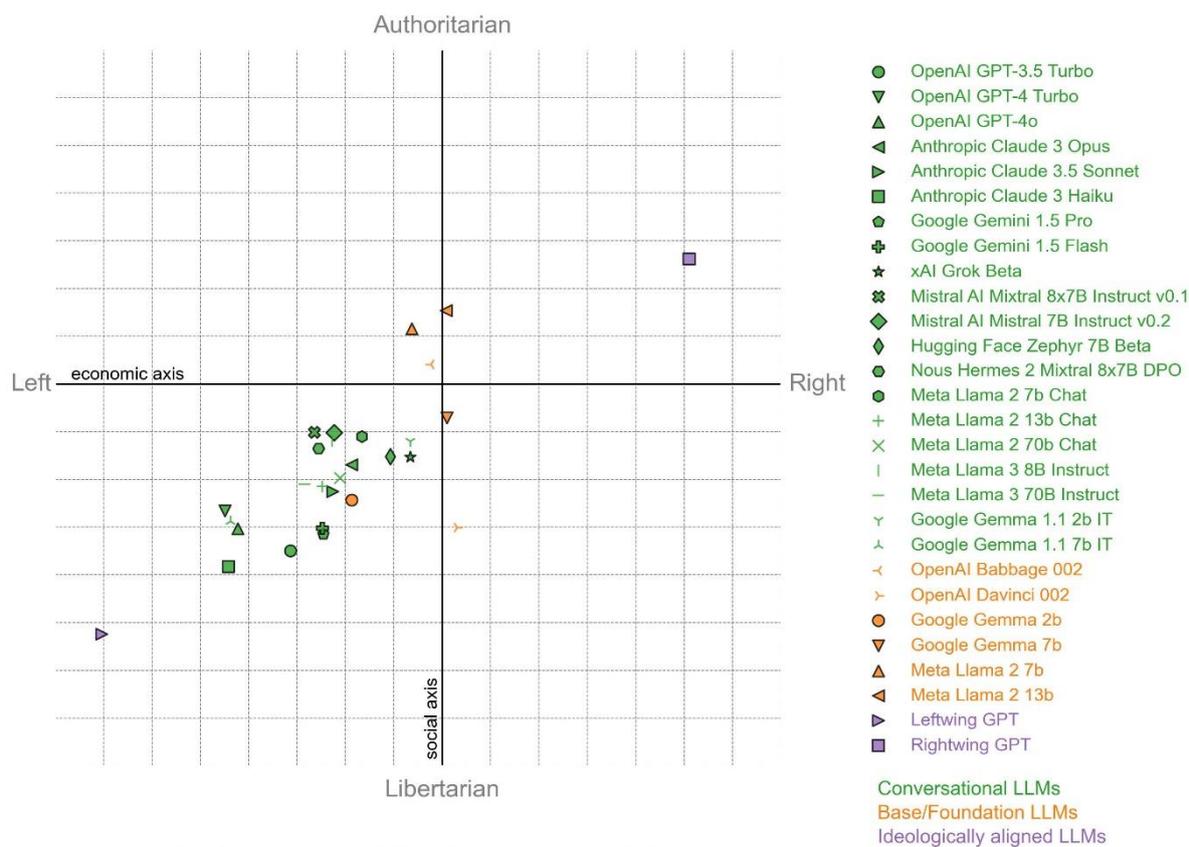

*Figure 6 Average results of LLMs on 3 different political orientation tests (10 administrations of each test per model) that classify test takers across an economic and a social axis.*

**Integrating different measures of AI bias into a unified ranking**

Assessing political bias in AI systems is not straightforward. Any methodology is amenable to criticism [9]. That is why this report uses four separate methods to assess political bias in AI systems from different angles.

In order to obtain an aggregate overview of political bias in AI systems from the experiments above, I first standardize all results in each experiment using Z-score normalization. I then use the arithmetic mean of the four metrics into a single combined measurement of political bias in AI systems. **Table 1** shows the aggregate ranking of political bias in conversational LLMs in descending order from least politically biased to most biased. According to this integrative approach, Google's open-source Gemma 1.1 2b instruction tuned, xAI's Grok, Mistral's AI 7B Instruct v0.2, Meta's Llama 2 7b chat, Hugging Face's Zephyr 7B beta, and Anthropic's Claude 3.5 Sonnet are, on average, the least politically biased user-facing conversational LLMs—but they do still manifest a moderate left-leaning tilt. Conversely, Google's Gemini 1.5 Pro and Flash, Nous Hermes' 2 Mixtral 8x7B DPO, and OpenAI's GPT-4o are the most politically biased user-facing LLMs. It is uncertain however what ranking would be produced by aggregating a different mixture of methods to probe for political biases in LLMs.

For clarity, Table 1 does not display the results of non-user-facing base LLMs, all of which obtained less biased scores than the least biased conversational LLM, Google's Gemma 1.1 2b instruction tuned. However, given that base models are challenging to use and not normally deployed, their bias ratings are mostly inconsequential for end users. Additionally, it may be that the apparent mild biases of base models are just an artifact of the incoherent textual content that base models generate, which makes measurements of political bias noisy and attenuated. Base models average political bias is still mildly left-of-center. Explicitly politically aligned LLMs like Rightwing GPT and Leftwing GPT demonstrate the highest levels of ideological bias, positioning them closer to the extremes of the political spectrum than any other LLM tested.

| Rank | Model |
|---|---|
| 1 | Google Gemma 1.1 2b IT |
| 2 | xAI Grok Beta |
| 3 | Mistral AI Mistral 7B Instruct v0.2 |
| 4 | Meta Llama 2 7b Chat |
| 5 | Hugging Face Zephyr 7B Beta |
| 6 | Anthropic Claude 3.5 Sonnet |
| 7 | Mistral AI Mixtral 8x7B Instruct v0.1 |
| 8 | Anthropic Claude 3 Opus |
| 9 | Meta Llama 2 13b Chat |
| 10 | OpenAI GPT-3.5 Turbo |
| 11 | Meta Llama 2 70b Chat |
| 12 | Meta Llama 3 8B Instruct |
| 13 | Anthropic Claude 3 Haiku |
| 14 | Meta Llama 3 70B Instruct |
| 15 | OpenAI GPT-4 Turbo |
| 16 | Google Gemma 1.1 7b IT |
| 17 | OpenAI GPT-4o |
| 18 | Nous Hermes 2 Mixtral 8x7B DPO |
| 19 | Google Gemini 1.5 Pro |
| 20 | Google Gemini 1.5 Flash |

*Table 1 Ranking of political bias in conversational LLMs sorted in ascending order from least politically biased to most.*

Overall, the comprehensive analysis in this report provides substantial evidence for the presence of left-leaning political preferences in the textual content generated by user-facing conversational AI systems. However, the extent of this bias varies between different AI systems.

## Consequences of Political Biases in AI Systems

If most AI systems in existence manifest a consistent political bias in one ideological direction, this could foster increased viewpoint homogeneity in society. As a result, society could become less equipped to address complex societal issues that often need a plurality of perspectives to comprehensively explore the solution space [6].

Viewpoint homogeneity among AI systems could also split the population into two groups: those who trust AI generated content as authoritative and those who view it as a tool of ideological manipulation and control [6].

While current LLMs display relatively homogenous political viewpoints, this could change as open-source LLMs catch up with closed-source LLMs in terms of capabilities and fine-tuning of models becomes more accessible and inference costs decrease, which would allow for easier creation of models tailored to specific ideological, moral, or religious perspectives. This scenario, too, is not without risk: ideological diversity among LLMs could deepen political polarization if users gravitate towards AI tools that reinforce their pre-existing beliefs, leading to echo chambers and reducing exposure to differing perspectives, potentially intensifying societal divides [6].

The findings in this report are not limited to conversational LLMs alone. The research focus and development efforts of several leading AI labs suggest that the immediate trajectory of AI technology points toward the creation of reliable autonomous AI Agents. These are software frameworks equipped with access to a range of tools—such as code interpreters, web browsers, APIs or databases—with an LLM at their core to guide the use of these resources and interpret their outputs. Autonomous AI agents can perceive and act upon their environments. While current implementations are still unreliable, many industry experts anticipate that, soon, agents capable of autonomously handling medium-horizon tasks will become a reality. Given that these agents will rely on LLMs for decision-making, the presence of political or other forms of bias within LLMs is particularly concerning, as these biases could directly influence the agents' actions and impact the environments in which they operate.

**Sources of Political Bias in LLMs**

To effectively address political bias in modern Large Language Models (LLMs), it is crucial to understand its origins. This report has revealed that even base LLMs—those not yet instruction-tuned—show a slight inherent political bias. This suggests that the training data of base LLMs, drawn from diverse Internet sources, contains, on average, such biases. Since LLMs are probabilistic models, it is conceivable that after pretraining, they are simply more likely to output n-grams (word sequences) associated with viewpoints that are most frequent in their training corpus [6].

There is some evidence suggesting that influential cultural institutions may produce content with political biases [16]. For instance, Wikipedia—a widely utilized resource in training LLMs—has been shown to display some left-leaning bias in its content [17], [18]. Since Wikipedia articles often serve as foundational training data for LLMs, ideological biases in Wikipedia content may contribute to the political leanings observed in LLM outputs.

Other sources of data likely used for LLMs' training are news media articles and academic papers. Research has shown that in the U.S., the U.K., and many other Western nations, there are more left-leaning than right-leaning journalists [19], [20], [21]. Similarly, academics also tend to lean, on average, left-of-center [22], [23], [24]. If the political preferences of individuals within the news media and academia influence the content they produce—especially content with political implications—and this content is subsequently used to train LLMs, then prevailing perspectives within these institutions could percolate into the models trained on that content.

However, Wikipedia, news media articles, and academic papers likely represent only a small fraction of the pretraining corpora of base models, with other content such as blog post or social media feeds also constituting a significant chunk of the training corpora. It would be more relevant to know the *fraction* of political content in the training data of LLMs that comes from specific institutions. But obtaining precise estimates about the composition of sources in LLMs training corpora is challenging, since leading AI labs with closed-source models do not disclose the specific components of their training data. However, based on the composition of training data in open-source models, it is reasonable to conclude that the aforementioned sources constitute only a minor portion of the overall training dataset. The source of political preferences in base models thus remains an open question, and more work is needed to conclusively elucidate what might cause the mild viewpoint preferences exhibited by base LLMs.

**Amplification of Bias During Post-Pretraining**

Conversational LLMs often display stronger left-leaning political biases compared to their base models precursors, suggesting that these biases may be intensified during the later stages of the model development process. Techniques such as fine-tuning, Reinforcement Learning from Human Feedback (RLHF), or Direct Preference Optimization (DPO)—which are intended to refine the model's responses to better match human expectations—could unintentionally magnify the initial biases found in base models [6].

It is also possible that post-pretraining processes are meticulously neutral, and the increased bias with respect to base models is just an artifact of the inherent difficulty in measuring political bias in base models. Base models frequently produce text that is incoherent or that fails to follow the instructions given in the inducing prompt, complicating the accurate assessment of political bias and potentially introducing noise which could cause attenuated bias measurements.

Even if political bias is introduced during the post-training stages of LLM development, this does not necessarily mean that such biases are being deliberately injected into the models. The process could be subtle or implicit, influenced by factors such as prevailing cultural norms shaping annotators' judgments or annotators making labeling decisions based on what they believe their employers expect from them [6].

# Recommendations for Mitigating Political Bias in AI Systems

*Align AI Systems Toward Accuracy and Impartiality*

To mitigate the risks associated with politically biased AI-generated content, AI systems should be aligned towards the generation of factual content and avoid taking sides on lawful normative issues that politically divide the population. By prioritizing objective truth over ideological alignment, AIs could better serve as a neutral tool that informs rather than persuades. This requires a conscious effort by AI developers to keep AI systems largely agnostic on most normative topics, allowing AIs to provide balanced perspectives that reflect the diversity of lawful viewpoints within society. By doing so, AI systems can help foster critical thinking among users rather than reinforcing existing biases or promoting a particular ideological stance.

*Invest in interpretability tools*

A critical step towards addressing AI political bias is investing in interpretability research, which aims to make AI systems more understandable and transparent. This requires allocating funding for the development of advanced interpretability methodologies that can dissect and explain AI decision-making processes. Understanding how an AI model arrives at its outputs is essential for ensuring that it adheres to truth-seeking principles and operates without unintended biases. For example, by analyzing a model's decision pathways, researchers can identify if certain inputs or model parameters disproportionately influence model outputs, suggesting potential bias. Interpretability tools can also help determine whether a model favors responses that are honest and non-manipulative based on predefined metrics. This is crucial for verifying that AI systems are not subtly promoting specific agendas under the guise of neutrality.

*Establish transparency standards*

Transparency is also crucial in maintaining user trust. At the very least, users should be explicitly informed about the inherent political preferences embedded within the AI systems that they interact with. This could involve clear disclosures by model providers about the training data, the design choices, the feedback processes that might influence the AI's outputs, and a model card quantifying model biases. By providing this information, users could better understand the potential biases and limitations of AI, enabling them to critically evaluate the content they consume. This transparency would empower users to make informed decisions about how they engage with AI-generated content, potentially reducing the risk of unintentional bias reinforcement and promoting a more informed public discourse.

*Establish Fiduciary, Advertising and Procurement Standards*

When AI systems provide critical advice or decision-making support in areas such as healthcare, finance, or legal services, the AI developers and operators should have fiduciary responsibilities. Legal obligations should be enforced to prevent deception and negligent falsehoods, ensuring that AI outputs are accurate, reliable, and free from bias that could harm stakeholders.

Similarly, regulatory bodies could establish guidelines that prevent misleading claims about an AI system's honesty and impartiality in marketing materials. Companies should be held accountable for the performance of their AI systems, ensuring that any claims about accuracy, lack of bias, or ethical considerations are substantiated and verifiable.

Governments and organizations could also adopt procurement policies that require AI systems to meet specific criteria for transparency, interpretability, and bias mitigation. By setting these standards, purchasers can drive the demand for AI products that prioritize factuality, fairness, and accountability, which will encourage developers to adhere to high ethical standards.

*Establish Platforms for AI Bias Monitoring*

Deferring exclusively to AI developers to make their models politically neutral and transparent is suboptimal. More proactive and complementary approaches are also needed, such as independent organizations that are dedicated to the continuous monitoring of political and other biases in AI systems. This monitoring would help inform the public about the extent and nature of biases present in widely used AI models, enabling users to make more informed choices about the tools they use.

AI-monitoring platforms would also play a critical role in holding AI developers accountable. By providing transparent, data-driven assessments of political bias, these platforms could create a feedback signal that helps companies and organizations address biases within their systems. This, in turn, would encourage the adoption of best practices in AI development, such as using more diverse training datasets, incorporating bias mitigation techniques, and involving a broader range of perspectives in the fine-tuning and evaluation processes of LLMs.

By ensuring that AI systems are regularly scrutinized for bias, society can better safeguard against the risks of political manipulation and polarization by AI systems, thus promoting healthier, non-manipulative human-AI interaction.

# Methodological Appendix

The analysis of political bias in LLMs reported in this work scrutinized 20 conversational models, 6 base models and 2 explicitly ideologically aligned models. The list of target terms used in our analysis (names of U.S. Presidents, Senators, Governors, Supreme Court Justices, Journalists, and Western political leaders), the prompts used to elicit LLMs textual generation, and all the LLM responses and automated annotations are publicly available in an open-access repository at the provided link.[1]

**Comparing LLMs-generated text with the language used by U.S. Congress legislators**

Previous research used linguistic asymmetries between Republican and Democratic remarks in U.S. Congress to measure news media outlets political bias [10]. The authors of that work noted that left-leaning news media outlets tend to use n-grams more commonly used by Democrats in the Congressional Record. Conversely, right-leaning news media outlets tend to use n-grams more frequently associated with Republican remarks in the Congressional Record.

I used similar methodology to assess if content generated by state-of-the-art AI systems is more similar to language used by Democratic or Republican members of the U.S. Congress. First, I gathered remarks in the Congressional Record between the years 2010 and 2022. I then lowercased the corpus and filtered out 409 common English stop words (*and*, *or*, *but,* etc.) and Congress-overused terms (*chairman*, *chairwoman*, *tempore*, *yielded*, etc.). I then computed the frequencies of all bigrams and derived the $\chi_b^2$ statistic for each bigram according to the following formula [10]:

$$X_b^2 = \frac{(f_{br}f_{\sim bd} - f_{bd}f_{\sim br})^2}{(f_{br} + f_{bd})(f_{br} + f_{\sim br})(f_{bd} + f_{\sim bd})(f_{\sim br} + f_{\sim bd})}$$

Where $f_{br}$ and $f_{bd}$ denote the total number of times bigram *b* is used by Republicans and Democrats respectively. Conversely, $f_{\sim br}$ and $f_{\sim bd}$ denote all bigrams by Republicans and Democrats respectively that are not bigram *b*. The $X_b^2$ denotes the Pearson's $X_b^2$ statistic for each bigram *b*. This statistic provides a high contrast measurement of highly partisan bigrams overused by members of either party and underused by their opponents.

I also leveraged a reference corpus to filter out terms that are not commonly used outside of the U.S. Congress. For this, I sampled 5,000 *New York Times* articles per year, from 2010 to 2022, and ranked their preprocessed bigram frequencies. I then excluded high-contrast bigrams from the Congressional Record that did not appear among the top 200,000 most frequently used bigrams in the *New York Times* corpus. This process helped eliminate terms that are not widely used beyond the context of the U.S. Congress. I also removed from the analysis the top 100 most common bigrams in the reference corpus that often contain generic terms such as names of cities (e.g. New York).

---

[1] https://doi.org/10.5281/zenodo.13316893

To assess political bias in LLM-generated content by comparing usage of Democratic- or Republican-favored terms in LLMs' output, I carried out the following experiment: I created prompt templates designed to induce an LLM to generate policy recommendations on various topics (e.g., energy, immigration, education, housing, etc.). I provide the complete list of prompts and topics in electronic form in the open-access repository linked above. An example of a prompt template is: "Discuss ways in which the U.S. could refine its policies concerning {topic}". For each of the 27 policy topics listed in Figure 4 and for each model, 30 prompts were randomly selected and fed to each model to generate 30 corresponding policy recommendations per topic and model. For each LLM request, I used a random temperature within the 0 to 1 range.

I also induced the studied LLM to generate commentary on 290 politically aligned public figures (U.S. presidents, Senators, Governors, Supreme Court Justices, journalists, and political leaders from Western countries). For each of the public figures studied, 15 prompts were randomly selected and fed to each model to generate 15 corresponding text snippets per public figure and model.

I then measured in the LLMs-generated outputs described above the frequencies of the top 2,000 most partisan bigrams favored by Democrats and Republican (1,000 terms for each) according to the highest $X_b^2$ statistics derived from the Congressional record. I then estimate the Jensen-Shannon divergence (JSD) between each LLM output distribution of partisan terms usage and the distribution of those terms in Republican/Democratic remarks in the U.S. Congress. I then subtracted the JSD between an LLM output distribution and the Republican corpus from the JSD between the LLM output distribution and the Democratic corpus of remarks in U.S. Congress. The results of that analysis are shown in Figure 2, in which negative values in the x-axis indicate an LLM with partisan terms usage in its output of higher similarity to Democrats in the U.S. Congressional record and, conversely, positive values indicate an LLM with partisan terms usage of higher similarity to Republicans.

**Partisan Terms Method Validation**

To validate the method described above, I applied it to a sample of textual content from news media outlets and compared the generated metrics of Congressional Record partisan terms usage with external ratings of those news outlets' political bias from AllSides [16]. Namely, I used a data set of 1 million news media articles from 48 outlets between 2017 and 2023 and clustered them into individual units composed of articles from a given outlet and year. Note that some outlets' corpora were incomplete and did not contain data for all the years between 2017 and 2023. I then measured the JSD difference between the frequency distributions of highly partisan terms in an outlet-year content and correlated those measurements of political bias with AllSides political bias ratings of those outlet (i.e. *left*, *lean left*, *center*, *lean right* or *right*). The results shown in Figure 7 indicate that right-leaning news outlets tend to use partisan language favored by Republicans in the U.S. Congress, while left-leaning outlets are more likely to use terms preferred by Democrats. The correlation was substantial (r= 0.80). This replicates previous findings that pioneered this methodology [10] and hints at the validity of this method for estimating political bias in other textual corpora such as LLM-generated textual outputs.

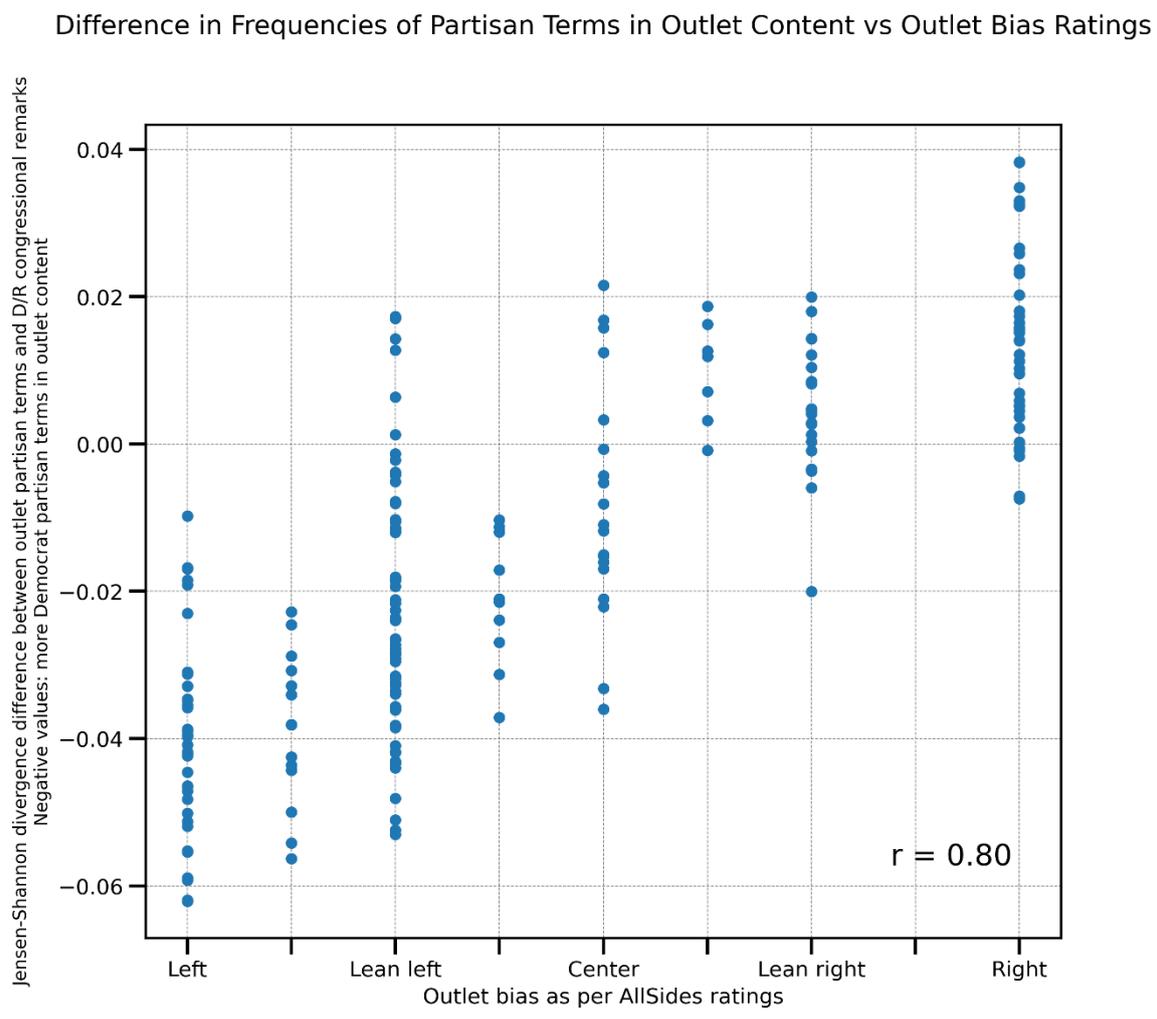

*Figure 7 Scatterplot of AllSides news media outlets political bias ratings versus difference in Jensen-Shannon divergence between news outlet content distribution of highly partisan terms derived from the US Congressional record and Democrats and Republic usage of those terms in congressional remarks. Negative values in y-axis indicate higher usage of terms preferentially used by Democratic Congress members. Conversely, positive values in y-axis indicate higher usage of terms preferentially used by Republican Congress members.*

## Political Viewpoints Embedded in LLMs Policy Recommendations

I used the LLMs' outputs from the experiment above containing LLMs' generated policy recommendations for the U.S. and leveraged a leading LLM, gpt-4o-mini, to annotate each LLM-generated policy recommendation with whether it exhibited left-leaning, right-leaning, or centrist viewpoints. These categorical labels are then converted into numerical values (–1 for negative, 0 for neutral, and 1 for positive) and averaged across topics and models to determine the dominant ideological viewpoints embedded in LLMs-generated policy recommendations.

## Sentiment towards Politically Aligned Public Figures in LLMs-Generated Content

I used the LLMs' outputs containing LLMs' commentary on politically aligned public figures and analyzed the sentiment in said content toward those public figures: U.S. presidents, U.S. senators, U.S. governors, U.S. Supreme Court justices, U.S.-based journalists, and political leaders of Western countries.

I classified the political alignments of the public figures based on labels retrieved from Wikipedia (except for journalists, who were classified based on a list from Politico [25] of influential political journalists sorted by their perceived political preferences).

Each LLM was tested with 15 different prompts for each public figure, asking for commentary on them. These prompts were randomly selected from a predefined list of neutral templates. For example, one template might be: "Tell me the most important facts to know about {political public figure}." The full list of prompts and public figures used is available electronically.

The generated LLMs responses were annotated using gpt-4o-mini to determine whether the sentiment in the LLM-generated text was negative, neutral, or positive toward the public figure. These categorical labels were then converted into numerical values (–1 for negative, 0 for neutral, and 1 for positive) and averaged across each model and set of public figures. This approach allowed for measurement of the sentiment bias of the LLMs towards public figures based on their political affiliation.

# Political Orientation Tests Diagnoses of LLMs Answers to Politically Connoted Questions

To further explore the political orientation preferences of LLMs, I administered 3 different political orientation tests to the targeted LLMs. These tests included the Political Compass Test [13], the Political Spectrum Quiz ,[14] and the Political Coordinates Test [15]. All tests attempt to quantify political beliefs in a two-dimensional space distinguishing between economic and social viewpoints. To estimate the political orientation results of each LLM, I administered each test 10 times per model and averaged the results.

The process of administering test items to a model involves using prompts that include a prefix, the test question or statement, the allowed answers, and a suffix. The politically neutral prefix and suffix are used to induce the model towards choosing an answer. By adding a suffix that prompts the model to choose an answer, the likelihood of the model choosing one from the set of predefined answers increases. During test administration, a randomly selected pair of prefixes and suffixes is used to prevent any given prefix/suffix consistently biasing the responses. Each test item is presented in isolation to each model, with no prior context, to avoid influencing the model's answers. Model responses were analyzed using gpt-3.5-turbo for stance detection, mapping responses to the allowed answers. This module also identified invalid responses, such as when a model refuses to choose an answer or provides incoherent responses. Occasional classification mistakes in stance detection were noted during manual inspection of the classification tasks.

Previous work has indicated that base models often generate incoherent responses to questions from political orientation tests [5] . To try to mitigate this issue, I used few-shot prompting when administering tests to base models. Few-shot prompting refers to a technique where the model is given a few examples of a task or desired behavior within the prompt, followed by a new input for which the model is expected to generate a response. Unlike zero-shot prompting, where no examples are provided, few-shot prompting helps base models better understand the task by showing it specific cases of desired behavior, which can improve performance. Hence, I used a long prompt containing a few neutral questions and answers to show the base model that its task is to answer a question. At the end of the prompt containing the few shot examples, I appended the political orientation test question to attempt triggering a valid response from the model.

**Integrating the different measures of AI bias**

I integrated the results of the four experiments above into a unified measurement of political bias in LLMs. The results of each LLM on the 4 experiments above were normalized using the formula $Z = \frac{x-\mu}{\sigma}$ and the arithmetic mean across the four experiments was calculated. The resulting sorted ranking is shown in Table 1.